\documentclass[10pt,aps,prd,amsmath,amssymb,twocolumn]{revtex4-1}

\usepackage[caption=false]{subfig} 
\usepackage{hyperref} 
\usepackage{graphicx,xcolor}
\usepackage{epstopdf} 
\usepackage{xcolor}

\def\pa{{\partial}}
\newcommand{\be}{\begin{equation}}
\newcommand{\ee}{\end{equation}}
\newcommand{\ba}{\begin{eqnarray}}
\newcommand{\ea}{\end{eqnarray}}
\newcommand{\nn}{\nonumber}

\newcommand{\h}[1]{\hat{#1}}

\newcommand{\hv}{\hbar v_F}

\begin{document} 
\title{Curvatronics with bilayer graphene in an effective $4D$ spacetime}

\author{Marco Cariglia}
\affiliation{Departamento de F\'isica, Universidade Federal de Ouro Preto, 35400-000 Ouro Preto MG, Brazil} 
\affiliation{School of Pharmacy, Physics Unit, Universit\`{a} di Camerino, 62032 - Camerino, Italy}
\affiliation{Dipartimento di Fisica e Astronomia, Universit\`a degli studi di Padova, via F. Marzolo 8, 35131 Padova, Italy}
 
\author{Roberto Giamb\`o}
\affiliation{School of Science and Technology, Mathematics Division, University of Camerino, 62032 - Camerino, Italy} 
 
\author{Andrea Perali}
\affiliation{School of Pharmacy, Physics Unit, University of Camerino, 62032 - Camerino, Italy}
\affiliation{INFN, Sezione di Perugia, 06123 - Perugia, Italy}


\begin{abstract} 
We show that in AB stacked bilayer graphene low energy excitations around the semimetallic points are described by massless, four dimensional Dirac fermions. There is an effective reconstruction of the 4 dimensional spacetime,  including in particular the dimension perpendicular to the sheet, that arises dynamically from the physical graphene sheet and the interactions experienced by the carriers. The effective spacetime is the Eisenhart-Duval lift of the dynamics experienced by Galilei invariant L\'evy-Leblond spin $\frac{1}{2}$ particles near the Dirac points. We find that changing the intrinsic curvature of the bilayer sheet induces a change in the energy level of the electronic bands, switching from a conducting regime for negative curvature to an insulating one when curvature is positive. In particular, curving graphene bilayers allows opening or closing the energy gap between conduction and valence bands, a key effect for electronic devices. 
Thus using curvature as a tunable parameter opens the way for the beginning of curvatronics in bilayer graphene. 
\\ 

\vspace{0\baselineskip} \noindent\textbf{Keywords}: 
Bilayer Graphene, L\'evy-Leblond equations, non-relativistic fermions, Eisenhart lift, curved systems. 
\\

\end{abstract}

\maketitle

\section{Introduction} 
The investigation of condensed matter systems, particularly in the nanometric regime and from the point of view of new materials with tunable electronic properties, has assumed a primary role in physics to understand many-body systems from first principles and to devise new technological applications. In the last few years a series of advances, both experimental and theoretical, have made evident that new materials in the quantum regime present a  range of phenomena that are well understood using concepts of relativistic particle physics, until recently thought to be removed from practical applications. Nowadays related activities are relevant topics in the physics community, and as an important example we cite that of studying massless and massive Dirac fermions in graphene. Here the state of the art is that of using the formalism of $3$D quantum field theory on curved spacetime to describe electronic properties of the material, where the effective curved geometry is originated by properties of the structure such as interaction with a substrate, topological defects such as disclinations and dislocations, or ripples \cite{Vozmediano2010,Yan2013,Ulstrup2014,Lewenstein2011,Szpak2014,Atanasov2011,Iorio2013,Cortijo2016,Beneventano2009}.
Distortions of the bilayer graphene lattice, induced for instance by an applied curvature, have been considered as a mechanism to tune and bend the quasiparticle energy dipersion, because they generate new phonons interacting with  conduction electrons in a non trivial way. Corresponding electronic and polaronic properties associated to these effects have been discussed extensively in Refs.\cite{LiEtAl2012,LiEtAl2011,LiEtAl2013,LiEtAl2015}.
In the study of superconductors, which also represent condensed matter systems with high potential for technological breakthroughs, there is some experimental evidence for fractal geometries in cuprates 
\cite{Buttner1987,Bianconi2010,Tranquada2004,Bianconi2011}, however no experiment in a curved geometry setting, while a recent theoretical work
investigates the effects of curvature on the superconducting pairing in the presence
of spin-orbit coupling, predicting non-trivial spin-triplet textures of the pairs \cite{YingEtAl2017}. In the case of carbon nanotubes effects of curvature on the electronic structure and transport have been studied \cite{Kleiner2001,Kane1997,Falk2010}. 
 
In this work we extend to bilayer graphene the relativistic approach and the methods of effective geometry for  the massless Dirac fermion of monolayer graphene, generalising it by showing that the geometry can include extra dimensions that are related to an effective reconstruction of the full ambient spacetime. We show that there is an effective 4D spacetime, in general curved, where solutions of the massless Dirac equation are in correspondence to low energy solutions of the original tight-binding model, in the continuum limit. This approach shows explicitly that even for bilayer graphene there are massless excitations, which is of interest on its own, and moreover is particularly powerful since  properties of the system can be inferred by known geometrical methods. For example we show in a simple way how the local 2D curvature affects the local energy density and the electronic structure. This is a kinematical effect that arises from the bound motion in a curved space, and is different in nature from the electrical gating effect, that is due to interaction with external fields. This provides a new mechanism to generate a gap in the energy levels. 

The outline of the paper is the following. 
In Section \ref{sec:free_LL} we begin showing that the quasi--free excitations in AB stacked bilayer graphene obey, in the low energy limit, the Galilei invariant L\'evy-Leblond equations for a spin $\frac{1}{2}$ particle. In Section \ref{sec:4D_free} we demonstrate that
 these solutions can be lifted to solutions of the massless Dirac equation in 4D Minkowski space. Next, we generalize this geometrical construction first in Section  \ref{sec:magnetic} by addition of a transverse, constant magnetic field, and then in Section \ref{sec:curved}, considering a curved 2D sheet of bilayer graphene and discussing its consequences on the electronic properties.  
In particular, the energy band gap is evaluated as a function of the curvature radius. Positive (resp. negative) curvature of the bilayer graphene is associated with insulating (resp. metallic) behavior of the system. Possible applications of {\sl curvatronics} are finally outlined in the concluding Section \ref{sec:conclusions}.

\section{Low energy solutions of the effective Hamiltonian in AB bilayer graphene\label{sec:free_LL}} The electronic band structure of graphite was studied in 1947 using a tight-binding model by Wallace \cite{Wallace1947}. Nowadays we know that for graphene there exist pairs of Dirac points $K$ and $K^\prime$ at the corners of the first Brillouin zone in momentum space, such that excitations with momentum sufficiently close to $K$ or $K^\prime$ display a linear dispersion relation, and we call these points 'valleys'. The distinct honeycomb lattice of graphene decomposes into two inequivalent $A$ and $B$ triangular lattices and these excitations are described by a massless, covariant, continuum theory for two degrees of freedom in two dimensions, obtained from the Dirac equation in the plane. The recent reference \cite{ZareniaPhD} contains technical details and a literature overview.

We start from the low energy limit tight-binding model Hamiltonian of AB stacked bilayer graphene close to the $K$ or $K'$ points  
\be \label{eq:H}
H_{K,K^\prime} = \left( \begin{array}{cccc} 
0 & \hv\kappa & 0 & \gamma \\ 
\hv\bar{\kappa} & 0 & 0 & 0 \\ 
0 & 0 & 0 & \hv \kappa \\ 
\gamma & 0 & \hv \bar{\kappa} & 0 
\end{array} \right) \, , 
\ee 
where $\kappa =  \tau k_x + i k_y $ is  the wave number of the excitation. $\tau = \pm 1$ denotes the Hamiltonian relative to the $K$ or $K'$ point,  $\gamma \sim 0.4 eV$ is the hopping parameter between $A_1$ and $B_2$ sites, while $v_F \sim 10^6 ms^{-1}$ is the Fermi velocity in a graphene monolayer close to the Dirac points.
 
The eigenvectors associated with the eigenvalue equation $H \lambda = E \lambda$ are  
\be \label{eq:sol_spinors}
\lambda_1 = 1 \, , \quad  \lambda_2 = \frac{\hv\bar{\kappa}}{E} \, , \quad \lambda_3 = \sigma \frac{\hv\kappa}{E} \, , \quad \lambda_4 =  \sigma \, ,  
\ee 
where $\bar{\kappa}$ is the complex conjugate of $\kappa$, $\lambda_j (j=1,\ldots,4)$ are the components of the eigenvectors,  and the energy $E$ satisfies the consistency condition 
\be \label{eq:sol_energy}
 E^2 - |\hv\kappa|^2  = \sigma \gamma E \, , \qquad \sigma = \pm 1 \, . 
\ee 
For each value of $\tau$ there are 4 solutions: for $\sigma = \pm 1$ there are two eigenvalues of the energy. The spectrum is valley degenerate, while the spinors \eqref{eq:sol_spinors} are not. We group the eigenvalues in two families:  
\be \label{eq:energy_free1}
E_i^{(\pm)} = \pm \left[ (-1)^{i} \frac{\gamma}{2} + \sqrt{ \frac{\gamma^2}{4} + |\hv\kappa|^2} \right] \, , i = 1,2. 
\ee  
The $E_1^{(\pm)}$ bands touch at $|\kappa|=0$ and make bilayer graphene a semi-metal, while the $E_2^{(\pm)}$ bands are separated by a distance $2 \gamma$. 
 
For low values of $\kappa$ all bands grow quadratically, which indicates non-relativistic behavior. In fact as we show in the rest of this section the low energy excitations satisfy the L\'evy-Leblond equations: non-relativistic, Galilei invariant equations for a spin $\frac{1}{2}$ particle of mass $m$ \cite{LevyLeblond1967}. These were written in 1967 as a non-relativistic limit of the Dirac equation, and proved that the $g=2$ Land\'e factor for the electron is not a relativistic property. In our case the mass is proportional to the hopping parameter $\gamma$.

We expand the solutions to leading order in the dimensionless parameter $\epsilon= \frac{\hv|\kappa|}{\gamma}$. For the $E_1^{(\pm)}$ bands 
\be \label{eq:expansion1}
\lambda_2 = \pm \frac{\gamma}{\hv\kappa} + O \left( \epsilon\right)  \, , \qquad \lambda_3 = \pm\sigma  \frac{\gamma}{\hv\bar{\kappa}} + O \left( \epsilon\right) \, , 
\ee
while for the $E_2^{(\pm)}$ bands 
\be \label{eq:expansion2} 
\lambda_2 = \pm \frac{\hv\bar{\kappa}}{\gamma} + O \left( \epsilon^3\right)  \, , \quad \lambda_3 =   \pm\sigma\frac{\hv\kappa}{\gamma} + O \left( \epsilon^3\right) \, .  
\ee  
The L\'evy-Leblond equations are written in terms of two time-dependent spinors $\chi_1(t)$, $\chi_2(t)$ with two components: 
\ba \label{eq:LevyLeblond1}
&&  i \hbar \, \pa_t \chi_2 + i \hbar v_F D \chi_1 = 0 \, , \\ 
&& D \chi_2 - i \frac{2 m v_F}{\hbar} \chi_1 = 0 \, , \label{eq:LevyLeblond2}
\ea 
where $D = i \sigma^j k_j$ is the $2$--dimensional Dirac operator in phase space and the $\sigma^j$ are the Pauli matrices in the standard basis. We replaced the speed of light $c$ in the original work with the relevant speed $v_F \sim \frac{c}{300}$ here.

\begin{figure}
\includegraphics[height=0.25\textwidth,width=0.45\textwidth]{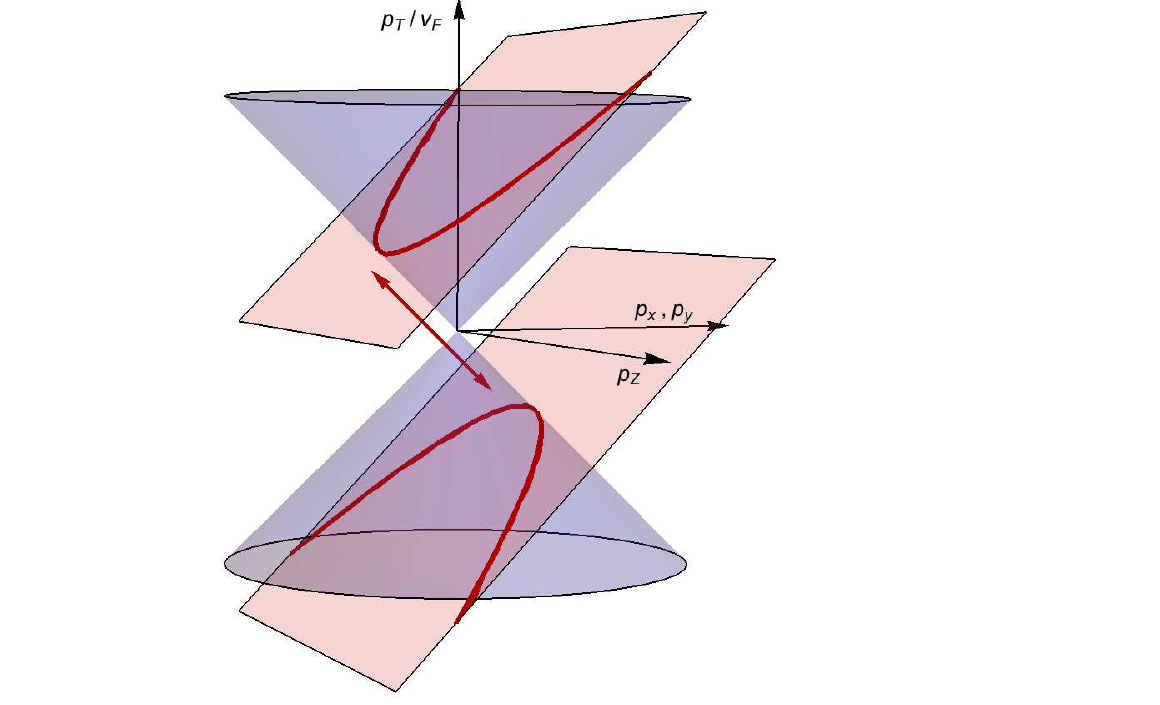} 
 \caption{\label{fig:cone}The 4D massless Dirac cone $p_x^2 + p_y^2 + 2 p_u p_v$ is cut by the plane $p_v = m v_F$. The result is the non-relativistic parabola $E = \frac{1}{2m}\left( p_x^2 + p_y^2 \right)$, where $E = - v_F p_u$. In our notation $p_Z = \frac{1}{\sqrt{2}}\left(p_u + p_v\right)$, $p_T = \frac{v_F}{\sqrt{2}}\left(p_u - p_v\right)$.}
  \end{figure}
 
Solutions of \eqref{eq:LevyLeblond1}, \eqref{eq:LevyLeblond2} can be obtained from \eqref{eq:expansion1} and \eqref{eq:expansion2}. For $E = E_1^{(\pm)}$ and $\tau = 1$ 
\ba 
\label{LL_solution_E1_tau+}
 \chi_1(t)  &=& 
 e^{\mp i \frac{|\hv\kappa|^2}{\hbar \gamma} t}  \left(  \lambda_1 , \sigma \lambda_4    \right)^T \, , \\ 
\chi_2(t)  &=& 
 e^{\mp i \frac{|\hv\kappa|^2}{\hbar \gamma} t}  \left(  \lambda_2 , \sigma \lambda_3   \right)^T \, , 
\ea 
where the mass is given by $m = \pm \frac{\gamma}{2 v_F^2}$.  For $E= E_2^{(\pm)}$ the solution is obtained swapping $\chi_1 \leftrightarrow \chi_2$ above, while for $\tau = - 1$ the solutions are generated by the substitution $\kappa \rightarrow - \bar{\kappa}$.  
Let us recall that $\lambda_1$ and $\lambda_4$ are related to components of the wavefunction on two stacked sites of type $A-B$ where up-down hopping is allowed, while the $\lambda_2$, $\lambda_3$ components are associated to sites which are not directly overlapping and for which hopping is negligible.

\section{Massless 4D fermions\label{sec:4D_free}} 
The massive L\'evy-Leblond equations can be obtained from the massless Dirac equation in a spacetime with 2 extra dimensions \cite{Duval1985,Duval1995_1,Duval1995_2,Cariglia2012}. The construction is based on the Eisenhart-Duval lift of dynamics, which was first discussed by Eisenhart \cite{Eisenhart1928} in the first half of the previous century, and then independently rediscovered by Duval and collaborators \cite{DBKP,DGH91}. The lift establishes a correspondence between classical, non-relativistic motions in the presence of a scalar and vector potential and null geodesics in a higher dimensional spacetime, and extends to quantum mechanics relating the non-relativistic Schr\"{o}dinger equation with the higher dimensional Klein-Gordon equation, and the L\'evy-Leblond with the Dirac equation.  The technique has been successfully used for several applications, as for example higher derivative systems \cite{Galajinsky2016} and non-relativistic holography \cite{Jensen2015,Geracie2015,Bekaert2016}, inspired by previous results in holography \cite{Zaanen2014}; see  \cite{CarigliaRMP2014} for a review of the associated geometry. 
 
As an example, the trajectory of a classical free point particle with unit mass moving on a plane can be lifted, using Eisenhart-Duval correspondence, to a null geodesic of the Minkowski metric in 4D written in double null coordinates $u$ and $v$: 
\be 
g_{\mu\nu} dx^\mu dx^\nu = dx^2 + dy^2 + 2 du dv \, .   
\ee 
The 4D variable $u$ corresponds, in a standard identification, with $ v_F t$, where $t$ is the time coordinate in the 2D dynamics. For a massless particle the null cone is $g^{\mu\nu} p_\mu p_\nu =0$, where $p_\mu$ are the momenta. Since the metric does not depend on $v$, then the condition $p_v = m v_F$  can be imposed. The intersection of this plane with the null cone yields the non-relativistic parabolic dispersion relation, as showed in Fig.\ref{fig:cone}. 
4D Gamma matrices $\Gamma^\mu$ satisfy the algebra $\{ \Gamma^\mu, \Gamma^\nu \} = 2 g^{\mu\nu}$. Spinors in 4D have 4 components, spinors in 2D have 2. We adopt a decomposition of Dirac matrices suitable for the null form of the metric: 
\ba  
\Gamma^+ &=&    \left( \begin{array}{cc} 0 & \mathbb{I} \\ 0 & 0 \end{array} \right)  , \,  \Gamma^- =   2 \left( \begin{array}{cc} 0 & 0 \\ \mathbb{I} & 0 \end{array} \right)  , \,  
\Gamma^i =   \left( \begin{array}{cc} \sigma_i & 0 \\ 0 & - \sigma_i \end{array} \right) \, . \nn 
\ea 
The 4D Dirac operator decomposes as 
\be \label{eq:4D_decomposed} 
\hat{D} =  \left( \begin{array}{cc} D &  \partial_u \\ 2 \partial_v & - D \end{array} \right) \, , 
\ee 
where $D = \sigma^i \partial_i$ is the 2D Dirac operator. The matrix above is not symmetric under the exchange of the $u$, $v$ coordinates. The factor of $2$ has been chosen in order to recover the L\'evy-Leblond equations, as discussed below.
Solutions of the massless Dirac equation in 4D, $\hat{D} \, \h{\Psi} = 0$, are compatible with the light-like projection 
\be \label{eq:projection}
\pa_v \h{\Psi} = i \frac{m v_F}{\hbar} \h{\Psi} \, ,  
\ee 
since the Minkowski metric is independent of $v$. Upon using this projection one immediately sees that \eqref{eq:4D_decomposed} induces the L\'evy-Leblond equations \eqref{eq:LevyLeblond1}, \eqref{eq:LevyLeblond2}. Therefore we reach the important conclusion that the 4D spinor  
\be \label{eq:spinor_4d_decomposition}
\h{\Psi} = e^{i \frac{m v_F}{\hbar} v}   \left( \begin{array}{c} \chi_1(u) \\ \chi_2(u) \end{array} \right) 
\ee 
satisfies the massless $4$--dimensional Dirac equation in flat space. The four possible spinors are in one to one correspondence with the low energy solutions of the tight-binding model of bilayer graphene close to the Dirac points. 
 
To conclude this section, we define 4D variables $Z, T$ using $u = \frac{Z +v_F T}{\sqrt{2}}$, $v = \frac{Z - v_F T}{\sqrt{2}}$, such that $2 du dv = dZ^2 - v_F^2 dT^2$. To lowest order in $\epsilon$ the $T, z$ dependence of our solutions is of the kind 
$e^{i \frac{\gamma}{\hbar v_F} \frac{Z - v_F T}{2\sqrt{2}}}$ 
from which we infer a wavelength along the $Z$ direction with value 
$ 
4 \sqrt{2} \pi \tfrac{\hbar v_F}{\gamma} = 29 \, nm, 
$ 
large compared to the real thickness of the bilayer. Therefore the effective description adopted here, where an effective flat space appears that is infinitely extended in all directions, is compatible with the real bilayer electronic structure: the wavefunction $\h{\Psi}$ is not able to resolve the real finite thickness.

\section{Transverse magnetic field\label{sec:magnetic}} The results obtained so far are not limited to the special case of free bilayer graphene. The only constraint is that the Eisenhart-Duval lift cannot describe external fields that depend on the $v$ direction: a dependence on the 2D time variable $t$ is allowed but not one on $Z$. So the case of an electric field transverse to the plane, and hence to the bilayer graphene, cannot be treated in this framework.  
 
In this section we show that our results continue being valid in the presence of a constant, transverse magnetic field $\vec{B} = B \, \hat{e}_Z$. The tight binding Hamiltonian is obtained from \eqref{eq:H} with the substitution 
\be
\hbar k \rightarrow \pi = - i \hbar \left( \tau \partial_x + i \pa_y + \frac{e B x}{\hbar} \right) \, , \ee 
arising from the standard definition of covariant momentum $\vec{p} - e \vec{A}$, and in a gauge where the only nonzero component of the gauge potential is $A_y = B x$. This Hamiltonian applies if the lattice spacing is much smaller than the magnetic length $l_B = \sqrt{\frac{\hbar}{eB}}$.  We look for solutions to the eigenvalue equation in the form $\lambda = e^{i k_y y} \varphi (x)$. In the rest of the section we will identify the operator $\pi$ with its reduction on the $\varphi$ type of spinors, i.e. we will write $\pi = - i \hbar \left( \tau \partial_x - k_y + \frac{e B x}{\hbar} \right)$. The problem reduces to the quantum harmonic oscillator since the rescaled operators $a =  \frac{1}{\sqrt{2\hbar e B} }  \pi = - \frac{i}{\sqrt{2}} \left(\tau \pa_\xi + \xi \right)$, and similarly for $a^\dagger$ satisfy the Heisenberg algebra. Here $\xi = \frac{x}{l_B} - k_y l_B$ is a dimensionless quantity. We employ the ansatz $\varphi_1 = \psi_n (\xi)$, where $\psi_n$ is the $n$--th level normalized eigenfunction of the quantum mechanical harmonic oscillator. For non-zero energy we find 
\be \label{eq:magnetic_energies_rescaled}
\tilde{E}_{1,2} = \pm \left( \frac{2n + 1 + \tilde{\gamma}^2 \pm \sqrt{\tilde{\gamma}^4 + (4 n + 2) \tilde{\gamma}^2 + 1}}{2} \right)^{\frac{1}{2}} \nn 
\ee  
where $\tilde{E} = \frac{E}{\sqrt{2\hbar e B}v_F}$ and similarly for $\gamma$, 
which agrees with the formulae reported in the literature, see e.g. \cite[Eqn. (2.49)]{ZareniaPhD}. In particular for the $n=0$ level there is no gap opening: this effect is the same in the case of the pseudo-magnetic fields arising from strain, when present. In the next section we will show that instead intrinsic curvature of the surface can open a gap: this underlies the difference between the effects of strain and those of curvature. 
The remaining spinor components are 
\ba 
&& \hspace{-0.6cm} \varphi_2 =  \frac{\sqrt{n+1}}{\tilde{E}} \psi_{n+1} \, , \, \varphi_3 = \sqrt{n} \, \frac{\tilde{E}^2 - (n+1)}{\tilde{E}^2 \tilde{\gamma}} \psi_{n-1} , \, \\ 
&& \hspace{-0.6cm} \varphi_4 =   \frac{\tilde{E}^2 - (n+1)}{\tilde{E} \tilde{\gamma}} \psi_{n} \, . 
\ea 
Now we examine the low energy limit and show that is is again described by the L\'evy-Leblond equations. 
For $B=0$ it must be $\frac{\pi \pi^\dagger + \pi^\dagger \pi}{2} =|\hbar \kappa|^2$, while for finite $B$ we have $\frac{\pi \pi^\dagger + \pi^\dagger \pi}{2} = 2 \hbar e B  (N + 1/2)$, where $N = a^\dagger a$ is the number operator. Therefore we take the limit $B \rightarrow 0$ together with $n \rightarrow + \infty$ so that $2 \hbar e B  v_F^2 (n + 1/2) \sim |\hv\kappa|^2 << \gamma^2$. In this limit the Landau levels of bilayer graphene become 
\be \label{eq:energy_magnetic1}
E_1^{(\pm)} \sim \pm \frac{2\hbar e B v_F^2 (n+1/2)}{\gamma} \, ,  \,   
E_2^{(\pm)} \sim \pm \gamma + E_1^{(\pm)}  \, ,   
\ee  
On the other hand the solutions of the L\'evy-Leblond equations for the $E_1^{(\pm)}$ bands are: 
\ba \label{LL_solution_E1_B}
\chi_1(t)  &=& 
  \left( \begin{array}{c} e^{\mp i n \frac{eB}{m} t} \, \frac{\sqrt{n 2 \hbar e B}}{2m v_F} \psi_{n} \\   e^{\mp i (n+1) \frac{eB}{m} t} \, \frac{\sqrt{(n+1) 2 \hbar e B}}{2m v_F} \psi_{n} \end{array} \right) \, , \\ 
\chi_2(t)  &=&   \left( e^{\mp i (n+1) \frac{eB}{m} t} \psi_{n+1} \,  , \,    e^{\mp i n \frac{eB}{m} t} \psi_{n-1} \right)^T 
\ea 
while for the $E_2^{(\pm)}$ bands 
\ba \label{LL_solution_E2_B}
 \begin{array}{c} \chi_1(t)  \end{array}  &=& 
  \left( \begin{array}{c}  e^{\mp i (n+1) \frac{eB}{m} t} \frac{\sqrt{(n+1) 2 \hbar e B}}{2m v_F} \psi_{n+1}  \\ e^{\mp i n \frac{eB}{m} t} \frac{\sqrt{n 2 \hbar e B}}{2m v_F} \psi_{n-1} \end{array} \right) \, ,  \\  
\chi_2(t) &=& \left( e^{\mp i n \frac{eB}{m} t} \psi_n \, , \,    e^{\mp i (n+1) \frac{eB}{m} t} \psi_n    \right)^T \, . 
\ea 
In the former case for low energy  $\left( \begin{array}{c} \chi_1(t) \\ \chi_2(t) \end{array} \right) \sim \frac{\sqrt{(n+\frac{1}{2}) 2 \hbar eB}}{2mv_F} e^{\mp i \frac{E_1^{(\pm)} }{\hbar \gamma} t}  \left(  \lambda_1 ,   \sigma \lambda_4,   \lambda_2 ,  \sigma \lambda_3   \right)^T$, and in the latter $ \left( \begin{array}{c} \chi_1(t) \\ \chi_2(t) \end{array} \right) \sim  e^{\mp i \frac{\left(E_2^{(\pm)} \mp \gamma\right)}{\hbar \gamma} t}  \left(   \lambda_2 ,   \sigma \lambda_3 ,   \lambda_1 ,  \sigma \lambda_4  \right)^T$, which are the same relations found in the free case. The mass of the low energy excitations is still given by $m = \pm \frac{\gamma}{2 v_F^2}$. Therefore, in this section we have demonstrated that the L\'evy-Leblond equations describe the low energy limit of the electronic spectrum, splitted in Landau levels, of bilayer graphene systems in the presence of an external magnetic field perpendicular to the layers.

\section{Extension to curved space: curvatronics\label{sec:curved}} 
To extend the example of the classical free point particle examined above, let us consider a generic 2D Riemannian space $\mathcal M$ with  metric 
$$
g=g_{ij}(x)\mathrm dx^i\mathrm dx^j,
$$ 
and the classical theory of a particle of mass $m$ and electric charge $e$ on $\mathcal M$, interacting with a scalar potential $V$ and an electromagnetic field with vector potential $A_i$, both possibly depending on position and time. In this case the Hamiltonian is given by  

$$
H=\frac{g^{ij}}{2m}(p_i- e A_i)(p_j- e A_j) +V \, .
$$
Then Eisenhart-Duval lift is given by the 4D Lorentzian metric 
\be 
ds^2 = g_{ij} \, dq^i dq^j + \frac{2e}{m v_F } A_i dq^i du + 2 du dv - \frac{2V}{m v_F^2}  du^2 \, , \nn 
\ee
where $q^i =(x, y)$, $A_i(q,u)$ and $V(q,u)$ are the vector and  scalar potential. Note that, the external potentials do not depend on the transverse $v$ direction, as pointed out above. To see that this is correct one can calculate the geodesic Hamiltonian from the metric above obtaining 
\be 
\mathcal{H} = \frac{g^{ij}}{2m}(p_i- \frac{e \, p_v}{mv_F} A_i)(p_j- \frac{e \, p_v}{mv_F} A_j) + \frac{p_u p_v}{m} + \frac{V}{m^2 v_F^2} p_v^2 \, . \nn 
\ee 
Setting $p_v = m v_F$, and $\mathcal{H} = 0$ for null geodesics we obtain the condition 
\be 
v_F \, p_u = - H \, . 
\ee 
If we define a new variable $t$ by $u = v_F \, t$ then the equation above says that $H$, the generator of time translations for the original dynamical system in 2D, can be identified with minus the momentum along the $t$ direction, which justifies identifying the variable $t$ with the time parameter in 2D.  
 
\begin{figure}%
         \centering
         \subfloat[$R<0$]{\includegraphics[height=0.1\textwidth,width=0.14\textwidth]{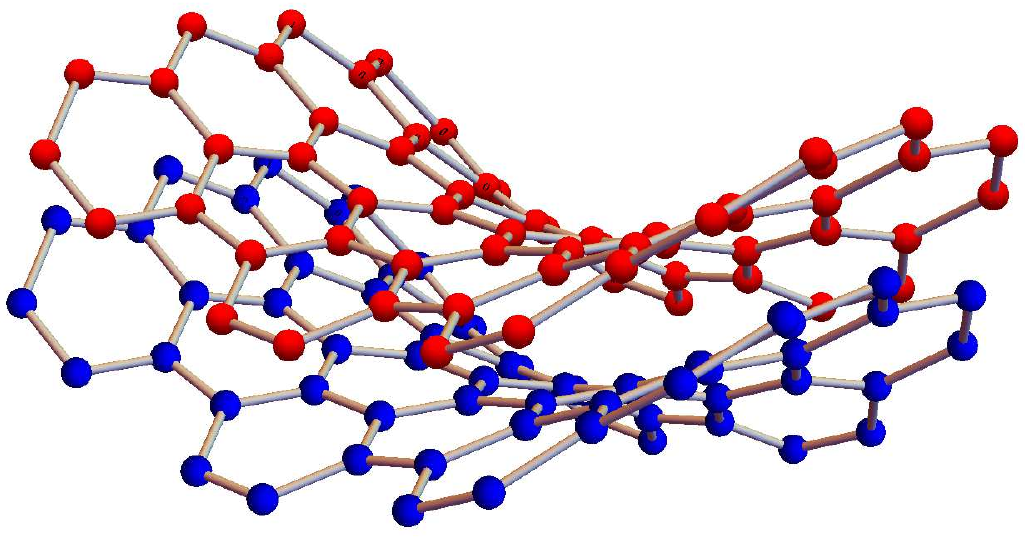}}\quad
         \subfloat[$R=0$]{\label{3figs-a}
         \includegraphics[height=0.08\textwidth,width=0.14\textwidth]{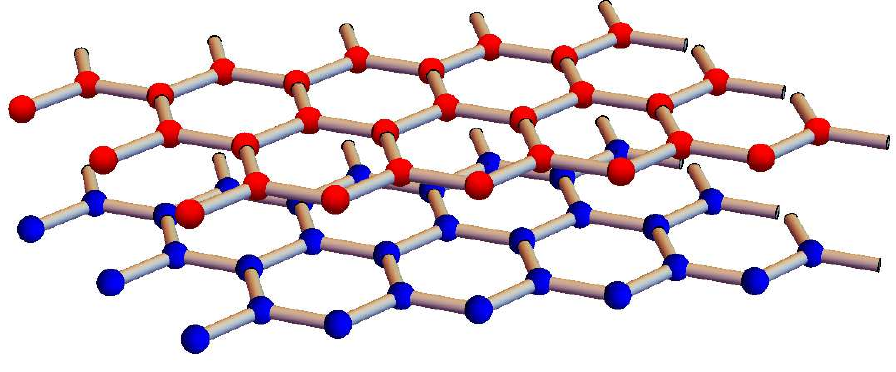}} \quad 
         \subfloat[$R>0$]{\includegraphics[height=0.1\textwidth,width=0.14\textwidth]{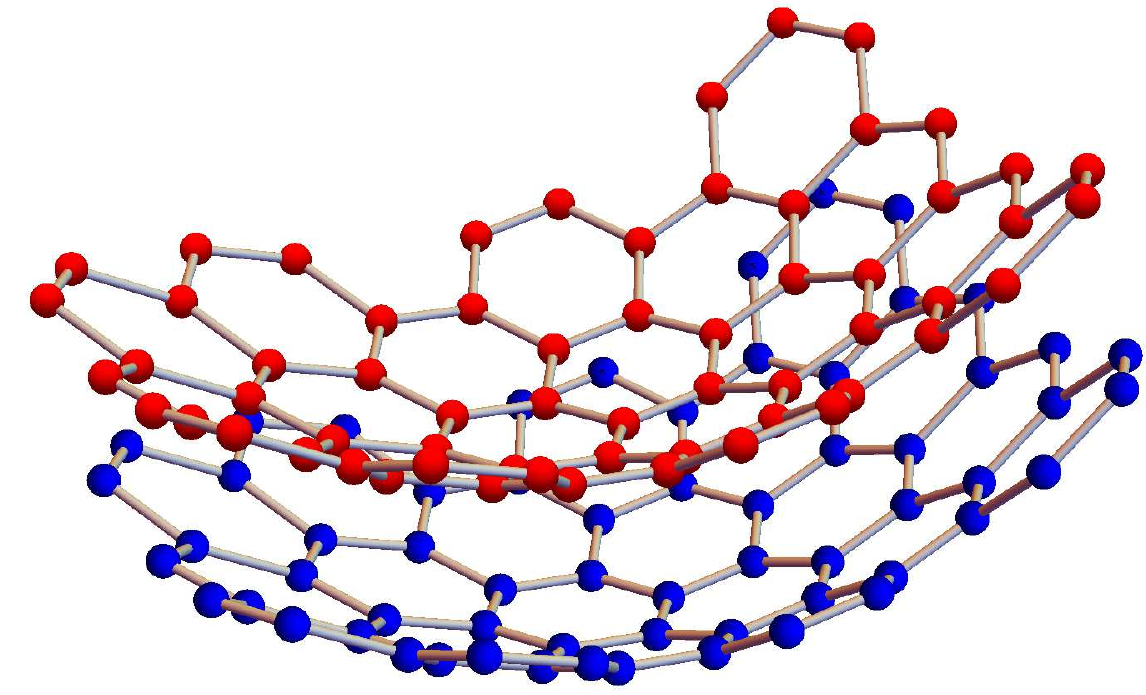}}
         \caption{Structures of bilayered graphene for different values of the 2D curvature. $R<0$ corresponds to hyperbolic geometry, $R=0$ to the flat graphene bilayer, $R>0$ to spherical geometry.}
         \label{3figs}
       \end{figure}

As a further generalization, $g$ can be considered as a time dependent 2D metric, which implies that $g_{ij}$ are also functions of $(q,u)$.  
Explicitly time dependent systems have been recently discussed in \cite{Cariglia2016}, where it has been shown that the formalism of the Eisenhart lift is compatible with the time dependence, both classically and quantum mechanically. To generalize the projection of the 4D Dirac operator 
\eqref{eq:4D_decomposed} described above to this case, the curvature of the Riemannian manifold $\mathcal M$
must be taken into account, and this can be accomplished using the covariant spinorial derivatives defined in \cite{Cariglia2012}, retrieving the operator -- with a slight correction with respect to \cite[Eqn. (4.11)]{Cariglia2012}
\be \label{eq:D_curved}
\hat{D} = \left( \begin{array}{cc} \frac{i}{\hbar} \, \Pi &  \partial_u + i \frac{V}{\hbar v_F} \\ 2 i \frac{m v_F}{\hbar} & - \frac{i}{\hbar} \, \Pi \end{array} \right) \, .
\ee 
Here $\Pi = \sigma^j \Pi_j$ and $\Pi_j =  - i \hbar \nabla_j - e A_j$ is the $U(1)$ covariant momentum including the spin connection, and we used the projection \eqref{eq:projection}. It can be seen from \eqref{eq:D_curved} that the massless, 4D Dirac equation $\hat{D} \hat{\Psi} = 0$ induces the curved version of the L\'evy-Leblond equations \eqref{eq:LevyLeblond1},\eqref{eq:LevyLeblond2} when $\hat{\Psi}$ is of the form \eqref{eq:spinor_4d_decomposition}. The equations obtained automatically include the presence of a scalar and vector potential, and of a curved 2D metric $g_{ij}$. In particular they incorporate the results above discussed for magnetic fields. Therefore non-relativistic L\'evy-Leblond fermions, arising in the continuum limit of the tight binding theory in bilayer graphene in presence of potentials independent of the variable $v$, are described by a massless Dirac equation in the effective 4D geometry. Bilayer sheets with constant curvature are represented in Fig.\ref{3figs}. In the literature on monolayer graphene it is known that, using a covariant version of the Dirac equation, one should consider effects of strain of the atomic lattice that produce pseudo-magnetic effective fields \cite{VozmedianoEtAl_inhomogeneities_2007,GuineaEtAl2010,VozmedianoEtAl_FermiVelocity_2012,VozmedianoEtAl_FromStrain_2013}. 
It is also known that the effects of strain are important, as they induce strong effective magnetic fields. The effective magnetic fields arising from strain modify the energy spectrum by inducing Landau levels. In particular, there always exists a zero Landau level and no gap in the energy spectrum can be opened in this way. As we are going to show next, our geometrical analysis shows that the local curvature of the sheet can be used to tune a gap opening: since this effect is different in nature from that of strain, in this work as a first step in the description of curved bilayer graphene we do not include the effects of strain. These are important and will be included in a future work.
The formalism we use has the important advantage of linking directly the curvature of the 2D geometry to the energy of states.  We realize the geometrical description with a locally Minkowski 4D metric, describing soft deformations of the bilayer structure that maintain the first order structure of the hexagonal cells, without local lattice strains or compressions of the bonds within the cells. For a non-trivial $g_{ij}$ the energy can be considered low if at all points $\hbar^2 v_F^2 R << \gamma^2$, where $R$ is the Ricci scalar of $g$. 

From Eq. \eqref{eq:D_curved} one obtains the curved version of the L\'evy-Leblond equations with a scalar potential. Taking two derivatives of $\chi_{2}$ we obtain the Schr\"{o}dinger equation 
\be \label{eq:Schrodinger_full} 
\left( \mathcal{E} - \frac{g^{ij}}{2m} \Pi_i \Pi_j - V + \frac{e \hbar}{2m} B  - \frac{\hbar^2}{8m} R \right) \chi_2 = 0 \, , 
\ee 
where $\mathcal{E} = i \hbar \, \pa_t$. 
\begin{figure}
\includegraphics[height=0.25\textwidth,width=0.45\textwidth]{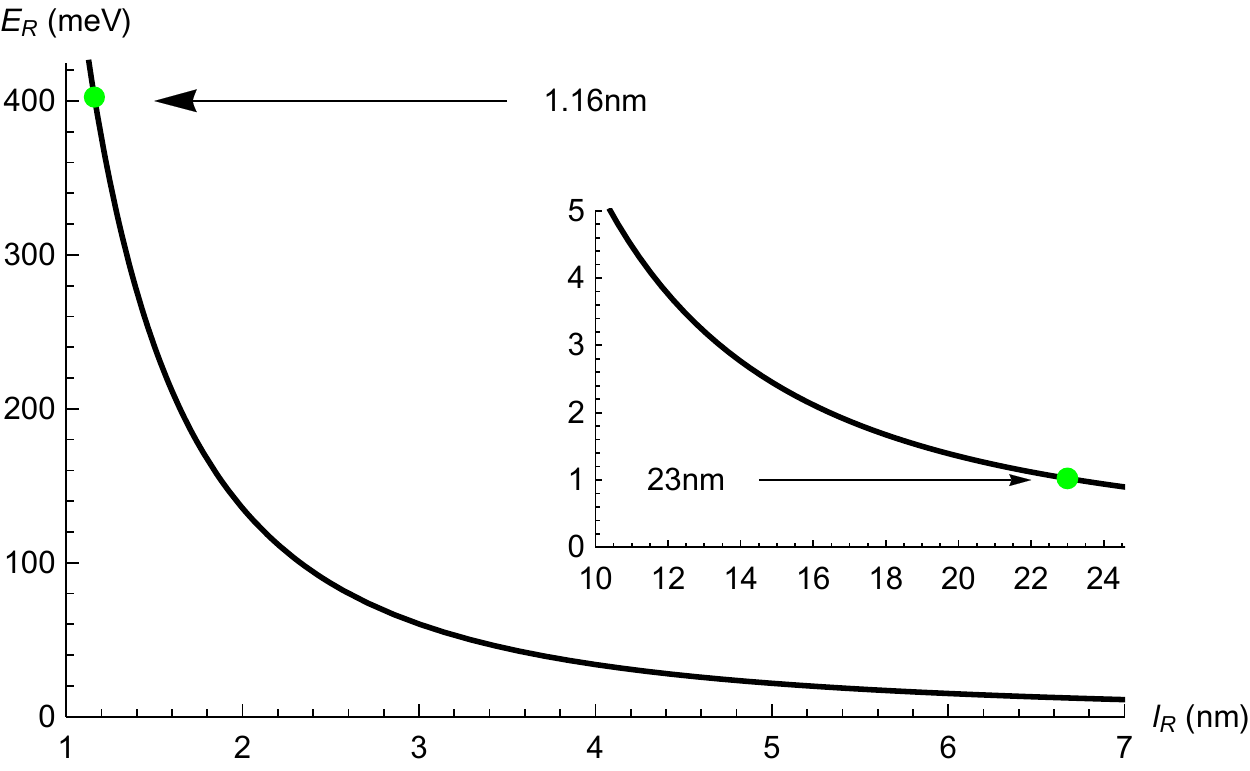} 
 \caption{\label{fig:bandgap}Energy band gap between conduction and valence bands $E_R$ as a function of the (constant) curvature radius $l_R$. The arrows indicate the validity range of our approach.}
  \end{figure}
For a surface of constant radius of curvature $l_R$ then $|R| = \frac{2}{l_R^2}$, and if the $R$ term in \eqref{eq:Schrodinger_full} is smaller than $\gamma$, we obtain $l_R >> 1.16 \,$nm, in agreement with requirements of smooth deformations on scales larger than the hexagonal cell. Experimental values of the energy contribution that can be measured by ARPES photoemission spectroscopy are of the order of $E_A = 10^{-3}eV$: requiring that the curvature effects are measurable with photoemission results in the constraint $l_R \le 23$nm. This is well within the typical curvature scale of interest in graphene systems \cite{NovoselovEtAl2007,CastroNetoEtAl2009,PeetersEtAl2014}. The scalar curvature of the surface has been discussed in the context of curved monolayer graphene in \cite{Lewenkopf2015}, where it has been associated to an effective pseudomagnetic field.

The case $V=0$, $B=0$ can be studied in terms of the eigenvalues $\mathfrak{p}$ of the spinorial momentum operator $\Pi$, proportional to the Dirac operator: then $\mathcal{E} = \frac{\mathfrak{p}^2}{2m}$. For example in the case of a sphere of radius $l_R$ the eigenvalues are known \cite{Varilly2006}, and the quantized energy is 
\be 
\mathcal{E} = \pm n^2 \frac{\hbar^2 v_F^2}{l_R^2 \gamma} \, \; \; n \in \mathbb{Z}^* \, , 
\ee 
where no zero-modes of the Dirac operator exist on the sphere. The expression is valid for $\frac{l_R}{n} > \frac{\hbar v_F}{\gamma} = 1.16$nm, and describes the two touching energy bands, as well as the departure from $\gamma$ of the non-touching bands. 
 
Our results imply that for a positive curvature surface the energy of $+$ bands will be shifted higher, while the energy of the $-$ bands will be shifted lower, due to the sign change in the effective mass. Therefore the shifted $E_1$ bands will make the bilayer graphene a semiconductor with a tunable band--gap proportional to $R$. In Fig. 
\ref{fig:bandgap} the energy band gap $E_R=E_1^{(+)}-E_1^{(-)}$ is plotted as a function of the curvature radius $l_R$, for positive constant curvature. Considering for instance the range $3$nm$<l_R<4$nm, Fig.\ref{fig:bandgap} shows that the band gap is in the range $E_R\simeq 40\div60 meV$, already enough to suppress thermal broadening and thermal excitations across the bands at or below room temperature. Band gap opening in bilayer graphene of energies of the order shown in Fig.\ref{fig:bandgap} have been obtained by electric field gating as measured in Ref.\cite{ZhangEtAl2009}, following the earlier prediction of Ref.\cite{MinEtAl2007}. On the other hand, negative curvature makes the material a conductor to leading order. Negative curvature can be applied to 2D semiconducting systems, as bilayer graphene with a band gap induced by an external potential, to reduce or close the band gap, increasing progressively their metallic behavior. 
 
The reader might wonder what is the physical reason behind the opening of a gap due to curvature, and if there is any relation with the previously known phenomenon of tunable gap opening by electrical gating. As remarked in \cite{ZhangEtAl2009}, a crucial reason why electrical gating opens a gap is that it breaks the inversion symmetry of bilayer graphene. We have investigated this issue and found that curvature does not break the inversion symmetry. Rather, while electrical gating is a dynamical effect, i.e. due to the interaction with an external field, the gap opening that we discuss in this work is kinematical in nature. The $R$ term present in the Schr\"{o}dinger equation \eqref{eq:Schrodinger_full} arises directly from the term 
\be 
\left[ \nabla_i , \nabla_j \right] = \frac{1}{4} R_{ijlm} \Gamma^{lm} 
\ee 
that expresses the non-commutation of (spinorial) covariant derivatives in curved space. In other words the gap arises from the properties of propagation of fields in the curved space, that are required by covariance and by consistency with the bound motion.

In the literature for monolayer graphene,  curvature is in connection with two opposite effects on electronic states: in an earlier work using a continuum model positive curvature was found to repel charges, and negative curvature to attract them \cite{AzevedoEtAl1998}, while more recent work that uses the Dirac equation on a curved background found that positive curvature conical defects are associated to an increase in the DOS, and viceversa for negative curvature  \cite{Vozmediano2007,Vozmediano2010}. On the one hand the results of \cite{AzevedoEtAl1998} should provide a refinement, smaller and in the opposite direction, of analyses based on the connectivity of single sites. On the other hand it is unclear if the results of \cite{Vozmediano2010} are due to the singularity or to the curvature: conical defects are singular points with zero intrinsic curvature, and the effects described are very localized, disappearing a few lattice constants away from the defect. Our approach describes long range curvature and therefore is complementary to that of \cite{Vozmediano2007}. In fact it is a second order effect: the energies in \cite{Vozmediano2010} are of the order of $E_{loc} = \frac{\hbar v_F}{l_R}$, while those in our model, using the value of $m$ in \eqref{eq:Schrodinger_full}, are $E_{geom} = \frac{\hbar^2 v_F^2}{2 \gamma l_R^2}$. In fact these are the two allowed combinations of parameters with which one can build an energy. Then our earlier requirement of low energy $\hbar^2 v_F^2 R << \gamma^2$ implies $E_{geom} << E_{loc}$. 
 
These results are important in the development of graphene based curvatronics as they give a powerful tool to describe the local effect of curvature on electronic states in \eqref{eq:Schrodinger_full}. They are also important in the fundamental understanding of bilayer graphene and can be applied to other 2D materials with massive quasiparticles. 

\section{Conclusions\label{sec:conclusions}} We have shown that the low energy limit of the continuum tight-binding model for AB stacked bilayer graphene is given by the Galilei invariant L\'evy-Leblond equations. Using the Eisenhart-Duval lift we proved that the low energy excitations satisfy the massless Dirac equation in an effective 4D Lorentzian geometry that reconstructs the full space. The parabolic dispersion relations of bilayer graphene look conical from a 4D perspective. 
We  presented detailed evidence for free bilayer graphene and bilayer graphene with a transverse, constant magnetic field. Application to a curved 2D sheet yields a simple and powerful relationship between the Ricci curvature of the surface and the local energy of the excitations, that arises from kinematical effects. The theory models the effect of long range curvature and is complementary, but with opposite behavior, to the theory of curvature generated by local defects. 
Our results open the way to curvatronics for tuning the electronic properties of graphene systems by local, smooth deformations, in such a way to allow and control a continuous crossover from metallic to semiconducting behavior and viceversa. Our geometrical approach can be also applied to other ultrathin materials and tested on naturally curved systems, as fullerens with their number of carbon atoms controlling the curvature, including fullerens with concentric onion-like structures having a spherical bilayer of carbon atoms generating a band gap \cite{Pincak2016}. Geometrical effects are also relevant for metamaterials with interesting topological properties, in which positive or negative curvature may induce very different effects and generate topological transitions \cite{KrishnaEtAl2012}.

\acknowledgments
We would like to thank L. Covaci, L. Dell'Anna, M. Doria, C. Duval, P. Horv\'athy, A. Marcelli, D. Neilson and M. Zarenia 
for useful discussions. 
M. Cariglia acknowledges CNPq support from project (205029/2014-0) and FAPEMIG support from project APQ-02164-14. 
A. Perali acknowledges financial support from 
the University of Camerino under the project FAR ``Control and enhancement of superconductivity by engineering materials at the nanoscale''. 
We acknowledge the collaboration within the MultiSuper International Network (http://www.multisuper.org) for exchange
of ideas and suggestions.


\vspace*{-1ex}

\begin{thebibliography}{99} 
 
\bibitem{Vozmediano2010} 
M. A. H. Vozmediano, M. I. Katsnelson, F. Guinea, \textit{Gauge fields in graphene}, Phys. Rep. \textbf{496}, 109 (2010). 
 
\bibitem{Yan2013} 
W. Yan et al., \textit{Strain and curvature induced evolution of electronic band structures in twisted graphene bilayer}, Nat. Comm. {\bf 4}, 2159 (2013).

\bibitem{Ulstrup2014} 
S. Ulstrup et al., \textit{Ultrafast Dynamics of Massive Dirac Fermions in Bilayer Graphene}, Phys. Rev. Lett. {\bf 112}, 257401 (2014).  
 
\bibitem{Lewenstein2011} 
O. Boada, A. Celi, J. I. Latorre, M. Lewenstein, \textit{Dirac equation for cold atoms in artificial
curved spacetimes}, New J. Phys. {\bf 13}, 035002 (2011). 
 
\bibitem{Szpak2014} 
N. Szpak, \textit{A Sheet of Graphene – Quantum Field in a Discrete Curved Space}, in Relativity and Gravitation, edited by J. Bi\v{c}\'ak and T. Ledvinka (Springer International Publishing, 2014) , p.583. 
 
\bibitem{Atanasov2011} 
V. Atanasov, A. Saxena, \textit{Electronic properties of corrugated graphene, the
Heisenberg principle and wormhole geometry in solid state}, J. Phys: Cond. Matt. \textbf{23}, 175301 (2011). 

\bibitem{Iorio2013} 
A. Iorio, \textit{Graphene: QFT in curved spacetimes close to
experiments}, J. Phys. Conf. Ser. {\bf 442}, 012056 (2013). 
 
\bibitem{Cortijo2016} 
A. Cortijo, M. A. Zubkov, \textit{Emergent gravity in the cubic tight-binding model of Weyl semimetal in the presence of elastic deformations}, Ann. Phys. {\bf 366}, 45 (2016). 
 
\bibitem{Beneventano2009} 
C. G. Beneventano, P. Giacconi, E. M. Santangelo, R. Soldati, \textit{Planar QED at finite temperature and density: Hall conductivity, Berry’s phases and minimal conductivity of graphene}, J. Phys. A \textbf{42}, 275401 (2009). 

\bibitem{LiEtAl2012}
Z. Li, L. Covaci, and F. Marsiglio,
\textit{Impact of Dresselhaus versus Rashba spin-orbit coupling on the Holstein polaron},
Phys. Rev. B \textbf{85}, 205112 (2012).

\bibitem{LiEtAl2011}
Z. Li, C. J. Chandler, and F. Marsiglio,
\textit{Perturbation theory of the mass enhancement for a polaron coupled to acoustic phonons},
Phys. Rev. B \textbf{83}, 045104 (2011)

\bibitem{LiEtAl2013}
Z. Li and J. P. Carbotte,
\textit{Conductivity of Dirac fermions with phonon-induced topological crossover},
Phys. Rev. B \textbf{88}, 195133 (2013)

\bibitem{LiEtAl2015}
Z. Li and J. P. Carbotte,
\textit{Electron-phonon correlations on spin texture of gapped helical Dirac fermions},
Eur. Phys. J. B  \textbf{88}, 87 (2015)

\bibitem{Buttner1987} 
H. B\"{u}ttner, A. Blumen, \textit{Possible explanation for the superconducting 240-K phase in the Y–Ba–Cu–O system}, Nature {\bf 329}, 700 (1987).  
 
\bibitem{Bianconi2010} 
M. Fratini, N. Poccia, A. Ricci, G. Campi, M. Burghammer, G. Aeppli, A. Bianconi, \textit{Scale-free structural organization of oxygen interstitials in La${}_2$CuO${}_{4+y}$}, Nature {\bf 466}, 841 (2010). 
 
\bibitem{Tranquada2004} 
J. M. Tranquada, H. Woo, T. G. Perring, H. Goka, G. D. Gu, G. Xu, M. Fujita, K.
Yamada, \textit{Quantum magnetic excitations from stripes in copper-oxide
superconductors}, Nature {\bf 429}, 534 (2004). 
 
\bibitem{Bianconi2011} 
N. Poccia, A. Ricci, A. Bianconi, \textit{Fractal structure favouring superconductivity at
high temperatures in a stack of membranes near a strain quantum critical
point}, J. Supercond. Nov. Magn. {\bf 24}, 1195 (2011). 
 
\bibitem{YingEtAl2017} Z.-J. Ying, M. Cuoco, C. Ortix, P. Gentile, \textit{Tuning Pairing Amplitude and Spin-Triplet Texture by Curving Superconducting Nanostructures}, \texttt{arXiv:1704.00578} (2017).

 
\bibitem{Kleiner2001} 
A. Kleiner, S. Eggert, \textit{Curvature, hybridization, and STM images of carbon nanotubes}, Phys. Rev. B. {\bf 64}, 113402 (2001).
 
\bibitem{Kane1997} 
C. L. Kane, E. J. Mele, \textit{Size, shape, and low energy electronic structure of carbon nanotubes}, Phys. Rev. Lett. {\bf 78}, 1932 (1997). 
 
\bibitem{Falk2010} 
K. Falk, F. Sedlmeier, L. Joly, R. R. Netz, L. Bocquet, \textit{Molecular origin of fast water transport in carbon nanotube membranes: superlubricity versus curvature dependent friction}, Nano lett. {\bf 10}, 4067 (2010).  

 
\bibitem{Wallace1947} 
P. R. Wallace, \textit{The band theory of graphite},  Phys. Rev. \textbf{71}, 622 (1947).
 
\bibitem{ZareniaPhD} 
M. Zarenia, \textit{Confined States in Mono-and Bi-layer Graphene Nanostructures}, Ph.D.  Thesis, Universiteit Antwerpen, 2013. 
 
\bibitem{LevyLeblond1967} 
J. M. L\'evy-Leblond, \textit{Nonrelativistic particles and Wave Equations}, Comm. Math. Phys. {\bf 6}, 286 (1967). 
 
\bibitem{Duval1985}
C. Duval, \textit{The Dirac \& L\'evy-Leblond Equations and Geometric Quantization}, 
in Differential Geometric Methods in Mathematical Physics, edited by P. L. Garc\'ia and A. P\'erez-Rendon  (Springer Berlin, Heidelberg, 1987), p.205
 
\bibitem{Duval1995_1}
C. Duval, P. A. Horv\'athy and L. Palla, \textit{Spinor vortices in nonrelativistic Chern-Simons theory}, Phys. Rev. D {\bf 52}, 4700 (1995). 

\bibitem{Duval1995_2}
C. Duval, P. A. Horv\'athy and L. Palla, \textit{Spinors in nonrelativistic Chern-Simons electrodynamics}, Ann. Phys.\  {\bf 249}, 265 (1996). 
 

\bibitem{Cariglia2012} 
M. Cariglia,  \textit{Hidden symmetries of Eisenhart-Duval lift metrics and the Dirac equation with flux}, Phys. Rev. D {\bf 86}, 084050 (2012). 
 
\bibitem{Eisenhart1928} 
L. P. Eisenhart, \textit{Dynamical trajectories and geodesics}, Ann. Math. \textbf{30}, 591 (1928). 
 
\bibitem{DBKP} 
C. Duval, G. Burdet, H. P. K\"{u}nzle and M. Perrin, \textit{Bargmann structures and Newton-Cartan theory}, Phys. Rev. D {\bf 31}, 1841 (1985). 

\bibitem{DGH91} 
C. Duval, G. W. Gibbons, P. Horv\'athy, \textit{Celestial mechanics, conformal structures and gravitational waves}, Phys. Rev. D {\bf 43}, 3907 (1991). 
 
\bibitem{Galajinsky2016} 
A. Galajinsky, I. Masterov, \textit{Eisenhart lift for higher derivative systems}, arXiv:1611.04294. 
 

  
\bibitem{Jensen2015} 
K. Jensen, A. Karch, \textit{Revisiting non-relativistic limits}, JHEP {\bf 2015.4}, 1 (2015).  
 

\bibitem{Geracie2015} 
M. Geracie, K. Prabhu, M. M. Roberts, \textit{Curved non-relativistic spacetimes, Newtonian gravitation and massive matter}, J. Math. Phys. {\bf 56}, 103505 (2015). 


\bibitem{Bekaert2016} 
X. Bekaert, K. Morand, \textit{Connections and dynamical trajectories in generalised Newton-Cartan gravity I. An intrinsic view}, J. Math. Phys. {\bf 57}, 022507 (2016). 
 
\bibitem{Zaanen2014} 
R. A. Davison, K. Schalm, J. Zaanen, \textit{Holographic duality and the resistivity of strange metals}, Phys. Rev. B {\bf 89}, 245116 (2014). 
 
\bibitem{CarigliaRMP2014} 
M. Cariglia, \textit{Hidden symmetries of dynamics in classical and quantum physics}, Rev. Mod. Phys. {\bf 86}, 1283 (2014). 
 
\bibitem{Cariglia2016} 
M. Cariglia, C. Duval, G. W. Gibbons, P. A. Horv\'athy, \textit{Eisenhart lifts and symmetries of time-dependent systems}, Ann. Phys. {\bf 373}, 631 (2016). 
 
\bibitem{VozmedianoEtAl_inhomogeneities_2007} 
F. de Juan, A. Cortijo, M. A. Vozmediano, \textit{Charge inhomogeneities due to smooth ripples in graphene sheets}, Phys. Rev. B {\bf 76}, 165409 (2007). 
 
\bibitem{GuineaEtAl2010}
F. Guinea, M. I. Katsnelson, A. K. Geim, \textit{Energy gaps and a zero-field quantum Hall effect in graphene by strain engineering}, Nat. Phys. {\bf 6}, 30 (2010). 
 
\bibitem{VozmedianoEtAl_FermiVelocity_2012}
F. de Juan, M. Sturla, M. A. Vozmediano, \textit{Space dependent Fermi velocity in strained graphene}, Phys. Rev. Lett. {\bf 108}, 227205 (2012). 
 
\bibitem{VozmedianoEtAl_FromStrain_2013} 
F. de Juan, J. L. Manes, M. A. Vozmediano, \textit{Gauge fields from strain in graphene}, Phys. Rev. B {\bf 87}, 165131 (2013). 
 
 
\bibitem{NovoselovEtAl2007} 
J. C. Meyer, A. K. Geim, M. I. Katsnelson, K. S. Novoselov, T. J. Booth, S. Roth, \textit{The structure of suspended graphene sheets}, Nat. {\bf 446}, 60 (2007). 
 
\bibitem{CastroNetoEtAl2009}  
A. C. Neto, F. Guinea, N. M. Peres, K. S. Novoselov, A. K. Geim, \textit{The electronic properties of graphene}, Rev. Mod. Phys. {\bf 14}, 109 (2009). 
 
\bibitem{PeetersEtAl2014} 
M. Neek-Amal, P. Xu, J. K. Schoelz, M. L. Ackerman, S. D. Barber, P. M. Thibado, A. Sadeghi, F. M. Peeters, \textit{Thermal mirror buckling in freestanding graphene locally controlled by scanning tunnelling microscopy}, Nat. Commun. {\bf 5}, 4962 (2014). 
   
 
\bibitem{Lewenkopf2015} 
E. Arias, A. R. Hern\'andez, C. Lewenkopf, \textit{Gauge fields in graphene with nonuniform elastic deformations: A quantum field theory approach}, Phys. Rev. B {\bf 92}, 245110 (2015). 


\bibitem{Varilly2006} 
J. C. V\'arilly, \textit{An Introduction to Noncommutative Geometry} (European Mathematical Society, 2006).  

\bibitem{MinEtAl2007}
H. Min, B. Sahu, S.K. Banerjee and A.H. MacDonald, 
\textit{Ab initio theory of gate induced gaps in graphene bilayers}, 
Phys. Rev. B \textbf{75} 155115 (2007). 

\bibitem{ZhangEtAl2009}
Y. Zhang,  T.-T. Tang, C. Girit, Z. Hao, M.C.  Martin, A. Zettl, M.F. Crommie, Y.R.  Shen, F.  Wang, 
\textit{Direct observation of a widely tunable bandgap in bilayer graphene}, Nature {\bf 459}, 820 (2009).

\bibitem{AzevedoEtAl1998}  
S. Azevedo, C. Furtado, F. Moraes, \textit{Charge localization around disclinations in monolayer graphite}, Phys. Stat. Sol.(b) {\bf 207}, 387 (1998). 

\bibitem{Vozmediano2007} 
A. Cortijo, M. A. Vozmediano, \textit{Electronic properties of curved graphene sheets}, Europhys. Lett. {\bf 77}, 47002 (2007).

\bibitem{Pincak2016} 
R. Pincak, V. V. Shunaev, J. Smotlacha, M. M. Slepchenkov, and O. E. Glukhova,
\textit{Electronic Properties of Bilayer Fullerene Onions}, arXiv:1612.01415

\bibitem{KrishnaEtAl2012}
H. N. S. Krishnamoorthy, Z. Jacob, E. Narimanov, I Kretzschmar, V. M. Menon,
\textit{Topological Transitions in Metamaterials},
Science {\bf 336}, 205 (2012).

\end{thebibliography}
\end{document}